\documentstyle[12pt,aaspp4]{article}
\def\IZw18{I~Zw~18}
\def\eqn{equation}
\def\ssp{\baselineskip=13pt plus 1pt minus 1pt}
\def\deg{\mbox {$^{\circ}$}}
\def\msun{\mbox {${\rm ~M_\odot}$}}
\def\zsun{\mbox {${\rm ~Z_{\odot}}$}}

\def\msunyr{\mbox {$~{\rm M_\odot}$~yr$^{-1}$}}
\def\angs{\mbox {~\AA}}
\def\Ha{\mbox {H$\alpha$~}}
\def\Hb{\mbox {H$\beta$~}}

\def\line{\mbox {~$\lambda$}}
\def\lines{\mbox {~$\lambda\lambda$~}}
\def\heha{\mbox {He~I~$\lambda 5876$~/~H$\alpha$}}
\def\xhe{\mbox {$\chi({\rm He}) / \chi({\rm H})$} }
\def\heii{\mbox {${\rm He}^+$}}
\def\he{\mbox {He}}
\def\hii{\mbox {${\rm H}^+$}}
\def\h{\mbox {H}}

\def\ohsun{\mbox {(O/H)$_{\odot}$~}}

\def\o3hb{[OIII]$\lambda5007$~/~H$\beta$~}
\def\O1ha{[OI]$\lambda6300$~/~H$\alpha$~}

\def\s2ha{[SII]$\lambda\lambda6717,31$~/~H$\alpha$~}
\def\2z2{HeII~$\lambda4686$~}
\def\z7{[NII]~$\lambda6583$ }
\def\N2{[NII]~$\lambda6583$ / H$\alpha$~}
\def\16z2{[SII]~$\lambda\lambda6717, 6731$ }

\def\n{NGC~}
\def\asec{\ifmmode {'' }\else $''~$\fi}  % arc sec
\def\amin{\ifmmode {' }\else $'~$\fi}    % arc min
\def\arcsper{\ifmmode \rlap.{'' }\else $\rlap{.}'' $\fi} % '' %Arcsec period
\def\arcmper{\ifmmode \rlap.{' }\else $\rlap{.}' $\fi} % '  %Arcmin period
\def\sles{\lower2pt\hbox{$\buildrel {\scriptstyle <}
   \over {\scriptstyle\sim}$}} % approximately less than
\def\sgreat{\lower2pt\hbox{$\buildrel {\scriptstyle >}
    \over {\scriptstyle\sim}$}} % approximately greater than
\def\gapp{\mbox {$_>\atop{^\sim}$}}  % approximately greater than
\def\kms{~km~s$^{-1}$~}

\def\sb{~ergs~s$^{-1}$~cm$^{-2}$~arcsec$^{-2}$}

\def\cm3{~cm$^{-3}$}

\def\fig{{Figure}}

\def\et{{\rm et\thinspace al.}\ }   % et al.
\def\apj{ApJ}
\def\apjs{ApJS}
\def\pasp{PASP}
\def\aj{AJ}
\def\mn{MNRAS}

\def\aa{A\&A}
\def\aasup{A\&AS}

\begin{document}

\title{Measurements of  He~I $\lambda 5876$ Recombination Line Radiation 
from the Diffuse, Warm Ionized Medium in Irregular Galaxies\altaffilmark{1}}

%\author{Crystal L. Martin\altaffilmark{1}
%and
%Robert C. Kennicutt, Jr.\altaffilmark{1}}

\author{Crystal L. Martin\altaffilmark{2,3,4}
and
Robert C. Kennicutt, Jr.\altaffilmark{2}}

\altaffiltext{1}{Observations reported here have been obtained in part with
the Multiple Mirror Telescope, a joint facility of the University of Arizona
and the Smithsonian Institution.}

\altaffiltext{2}{Steward Observatory, University of Arizona, Tucson, AZ 85721}

\altaffiltext{3}{Currently, Space Telescope Science Institute, 3700 San Martin
Drive, Baltimore, MD 21218}

\altaffiltext{4}{Hubble Fellow}

%ABSTRACT
\begin{abstract}
We present longslit optical spectroscopy of three high surface brightness
Magellanic irregular galaxies.
This paper draws attention to our detection of He~I\line 5876 line
emission  from the  ionized gas outside the HII regions, or the warm
ionized phase of the interstellar medium.  We measure 
a mean reddening-corrected intensity ratio of \heha\  $\approx 0.041$
independent of spatial location. This ratio is much higher than that
measured in the diffuse, warm ionized interstellar medium
of the Milky Way (Reynolds \& Tufte 1995).

%the high ionization fraction of He  implies 
%The presence of young, massive stars  (and clusters) in these galaxies
%is consistent with the hypothesis that the DIG is powered by stellar
%radiation.
%If the  ionizing continuum is produced by  massive stars,
%stars of mass $\ge 35 \msun\ $ must contribute to the maintenance of the DIG.
%Since
%optical and ultraviolet spectra confirm the presence of such massive
%stars in these galaxies, it seems quite likely that the DIG is powered
%by their radiation.

% $\xhe \approx 1$ 

% The high value of \heha\ implies the  \heii\ ionization fraction,
% $\chi({\he}) \equiv  n(\heii) / n(\he)$,
%  is approximately equal to the \hii\ ionization fraction,
% $\chi({\h}) \equiv n(\hii) / n(\h)$, in the diffuse ionizad gas (DIG).

%$\chi({\he}) \equiv  n(\heii) / n(\he),$

The high value of \heha\ implies the  helium ionization fraction
is approximately equal to the hydrogen ionization fraction
in the diffuse ionized gas (DIG). If the  DIG is powered by young stars, then
stars hotter than
% $\ge 35 \msun\ $ 
40,000~K must contribute to the Lyman continuum radiation reaching the DIG.  
Since optical and ultraviolet spectra confirm the presence of such massive
stars in these galaxies, stellar photoionization remains the most likely
power source.
The contrast with  the  low helium ionization in the Galactic DIG, however,
is intriguing and provides  strong 
evidence that the physical  state of the DIG, not just its presence, 
varies among galaxies.

%We speculate that fundamental differences in
%the spatial distribution of star formation, such as cluster size,  
%may explain the differences in ionization state.

\end{abstract}


\section{Introduction}
Warm, diffuse ionized gas (DIG) comprises a significant fraction of the mass 
and volume of the interstellar medium (ISM) in galaxies.  
In the Milky Way,  this component is sometimes referred to as the
Reynolds Layer and  has a scale height $\sim 1$~kpc (Reynolds 1991).
Galactic O and B stars supply enough power to keep it ionized
(Reynolds 1984), and  the only conundrum had been how the Lyman continuum
radiation from O and B stars, with scale height $\sim 100$~pc, can reach the DIG.
The  close correlation between the  presence of extended, diffuse ionized gas 
in other galaxies and various  tracers of massive star formation 
seemed to place  the OB-star photoionization hypothesis on relatively firm ground
(Hunter \& Gallagher 1990; Dettmar 1992; Rand 1996). 

The massive-star photoionization picture has been challenged, however, by new 
measurements of the He ionization fraction in the Galactic  DIG.
First, a search for  He I $\lambda 5876$ recombination
line emission from  the local DIG set an upper limit on the \heha\ intensity
ratio of $0.011$ (Reynolds \& Tufte 1995).
For a He/H abundance ratio of 1:10 by number in the DIG, 
the implied relative ionization fraction is
\xhe~$ \equiv$~ (n(He$^+$)/n(He))~$/$~(n(H$^+$)/n(H)) $ ~\sles~ 0.25$.
This high neutral fraction of He implies that the interstellar radiation field
is softer than that expected from the Galactic O star population in the
solar neighborhood (Reynolds \& Tufte 1984).
Further work has shown that this curiosity is not confined to the solar
neighborhood.  
Heiles \et (1996)  observed  hundreds of positions toward the Galactic center 
at $\sim 1.5$~GHz. The relative strengths of the
H and He radio recombination lines from the
quasi-vertical filaments of ionized gas known as ``worms''
indicate a relative  ionization fraction   $\xhe $  $~\sles~ 0.13$ there,
which implies that  stars more massive than about   39 \msun\ must not contribute
to the ionizing continuum (Heiles \et 1996).
To meet the measured ionization rate in the Galaxy, however, the global
star formation rate would need to be significantly  higher than
previously estimated (Heiles \et 1996).

%It remains unclear whether another excitation mechanism is involved,
%models  of the ionizing luminosity from Galactic stars are inaccurate,
%or the  radiative transport somehow softens the stellar  Lyman continuum.

Nearby galaxies provide an opportunity to examine
the relationship between the properties of the DIG and  
the stellar content of a galaxy.
Among normal galaxies, Magellanic irregular galaxies have the most intense
star formation in terms of both the number of HII regions per
unit luminosity and the  ionizing luminosity of the brightest HII region
(Kennicutt, Edgar, \& Hodge 1989).
To investigate the ionization state of He in the warm ionized phase of the 
ISM of such galaxies, we selected three irregular galaxies with 
copious amounts of  diffuse and filamentary \Ha emission
from a larger sample of dwarf galaxies (Martin 1996b).
In this paper, we report  measurements of the
He~I~$\lambda 5876$  line emission in the DIG from  longslit spectra.
A more detailed discussion of the emission-line spectra and their implications for the
excitation of the DIG can be found in Martin (1997a).

\section{Measurements of He~I\line 5876 Line Emission}

\subsection{The Galaxies}

The galaxies \n4214, \n1569, and \n4449 
were selected for their star formation activity,  proximity,
and prominence of  ionized gas beyond the HII
regions.  They are gas-rich  and have absolute blue luminosities between those 
of the  Small Magellanic Cloud and Large Magellanic Cloud. 
Table~1 compares some relevant properties of the observed galaxies to the
Milky Way.
Their current rates of massive star formation,  0.25 and 0.50 \msunyr,  
are only $\sim 5$
 times lower than that in the Milky Way but,  unlike our
Galaxy, are equal to or a few times larger than the past average rate 
of star formation in each galaxy 
(Kennicutt 1983;  Kennicutt, Tamblyn, \& Congdon 1994).
Spatial and temporal
fluctuations in the star formation rate  have clearly occurred.
For example, the northern region of \n4449 is younger than the main bar
(e.g. Hill \et 1994); and
\n1569 is emerging from a major burst of star formation 
(Israel \& de Bruyn 1988; Waller 1991; Heckman \et 1995).
Their lower oxygen abundance, O/H $\gapp ~0.20$~\ohsun, is also consistent with a 
much lower total amount of star formation in the past
(Martin 1997a).

\subsection{Observations and Reductions}

We obtained longslit spectra at 11 positions across these
galaxies at the Multiple Mirror Telescope during the period
February 1994 to March 1995.
The Blue Channel spectrograph was used with a 500~gpm
grating, a Loral 3k~$\times$~1k CCD detector, and a 1\asec by 3\amin slit.
This setup produced wavelength coverage from 3700 -- 6800 \angs\ and 
a spectral resolution of 4 - 5 \angs\ at \Ha.
Slit positions were chosen in advance to sample both HII regions and 
extended, diffuse emission and  were required to be close to the parallactic 
angle at the time of observation.
Their exact locations are illustrated in Martin (1997a), where additional 
details about the observations and reductions can be found.  The spectra reach
a surface brightness $\sim 2.5 \times 10^{-17}$\sb, which corresponds to an
emission measure of 14 at $T_e = 10^4$~K.

The  \ion{He}{1} \line 5876 line is clearly visible along a substantial length
of the slit in the raw data.  We divided this region into   3\asec apertures and
extracted  a series of one-dimensional spectra 
from the  sky-subtracted and continuum-subtracted frames as described by Martin
(1997a).  Fortunately, the recessional velocities of these galaxies are less
than 300\kms, so the line is cleanly separated from the bright 
night sky emission at \ion{Na}{1} \lines\ 5890,96. Our continuum template
does not include line emission, so stellar absorption/emission lines
are present in the continuum-subtracted frames.  The correction to the nebular
\ion{He}{1} \line 5876 emission will only be significant where the emission line
is weak relative to the continuum, and very little continuum emission underlies
most of the low surface brightness DIG in our spectra.  Measurements from
our spectra of the HII regions, however, do show a slight trend for the 
\heha\ ratio to decrease as the emission-line equivalent width decreases.
The slope of this relation places an upper limit of 0.3~\angs\ on the 
equivalent width of the underlying \ion{He}{1} \line\ 5876 absorption there.

The \ion{He}{1} \line 5876, \Ha, and \Hb  fluxes were measured from these spectra
using the `splot' task in IRAF.
The \heha\ ratio was corrected for reddening using the extinction curve of
Miller and Matthews (1972) and the logarithmic extinction at \Hb, $c(\Hb)$.
We derived  $c(\Hb)$ from the  Balmer decrement assuming a constant underlying 
stellar absorption equivalent width of $2\angs\ \pm 2\angs$ and 
an electron temperature of $15,000$~K.  
We found little variation of $c(\Hb)$ with
position along the slit, so the mean value was adopted at each position angle.
An uncertainty, $\delta c(\Hb)$  in Table~\ref{tab:table1},
was assigned based on the variation in $c(\Hb)$ along
the slit or the formal error, whichever was larger.  This term dominates the error
estimates of the \heha\ intensity ratio.  Only a portion of the data were obtained
under photometric conditions, but the relative flux calibration for all observations
is good to $\sim 2 \%$, based on observations of multiple standard stars
(Massey \et 1988).

%(e.g. Osterbrock 1989). 
%intensities of the \Ha and \Hb lines assuming the case~B Balmer decrement at

\subsection{Results}

Figure~1a shows the reddening-corrected ratio of the He~I\line 5876 to
\Ha emission-line intensity as a function of \Ha surface brightness
along  the 11 slit positions.
The \Ha surface brightness is closely correlated with the
angular distance from the nearest giant HII region  along our slit positions
(see \fig~\ref{fig:slit})
and  can therefore represent the relative distance of the aperture from the 
ionizing cluster.
The average \heha\ intensity ratio is 0.041,  the lowest
value is 0.028, and the highest value is 0.058.  
Comparison with panel b shows only \n1569 and position~3 in \n4449
(PA = 137.2\deg) have large corrections for reddening.
Across each  galaxy,  we find no systematic variation in \heha\
despite  a decline in \Ha\ surface brightness by a factor of  $\sim 100$.

The gradients in  other diagnostic line ratios measured from the 
same spectra emphasize the remarkable uniformity of \heha.
Figure~2 demonstrates the contrast along position~2 in \n1569.
Across the region where \heha\ is measured, the \s2ha, \N2, and \O1ha ratios
increase by factors of a few, while \o3hb decreases by  a similar factor.
This spectral change is typical of the DIG in low metallicity galaxies  (Martin 1997a).
Martin (1997a) studied these spectral changes using  photoionization models and found
the gradients in the line ratios primarily  reflect a gradient in  
the relative density of ionizing photons to gas. 
The ionization parameter is inferred to fall by a factor $\sgreat\ 10$ over the region
where we have measured a constant \heha\ ratio (Table~2).

% shows  similar examples across \n4449 and \n4861.

Under normal conditions, the relative intensity  of \heha\  can be
predicted from  the effective recombination coefficients of \he\ and \h.
At $T = 10^4$~K and $n = 100$\cm3, the recombination
coefficients for He and H from Brocklehurst (1972) and Hummer \& Storey (1987),
respectively, yield an emissivity ratio
	\begin{equation}
 \frac{E_{5876}}{E_{H\alpha}} = 0.470 \frac{\he}{\h} \frac{\chi(\he)}{\chi(\h)},  
	\end{equation}
where $\he / \h$ is the abundance ratio by number.
(The revised emissivities of Smits (1996) would raise the
coefficient in \eqn\ 1 by  0.004, while raising the temperature to
$1.2 \times 10^4$~K would  lower it by 0.004).
Assuming that the abundance ratio of  He / H by number is $\sles\ 0.1$
and comparing the  measured intensity ratios with \eqn\ 1, we see that
$\xhe \approx 1$.

If we adopt $\xhe = 1$, then the mean ionic abundance of \heii\ relative to \hii\
is $\frac{\heii}{\hii} \equiv \bar{y^+} = 0.085$, which is consistent
with the He/H ratio predicted by the He~vs~O regression relation of Pagel \et 
(1992) for the oxygen abundance,  $\log ({\rm O/H}) = -3.69 \pm 0.07$,
in  \n4449.  The slightly lower O/H ratios in
Table~1 for \n1569 and \n4214 are not unusual for an abundance ratio
He/H~$\approx 0.085$.  
For example, within the O/H range of our three irregular galaxies, 
Table~15 of Pagel \et (1992) contains HII regions with  $y^+$ varying 
from 0.081 to 0.090.
Also, Kobulnicky \& Skillman (1996) find variations in O/H
within \n4214 as large as $\sim 45\%$ whereas He$^+$/H$^+$ changes by only
$\sim 6\%$ among the same regions.

\section{Discussion:  The Source of Ionizing Photons}

\begin{figure}
\plotone{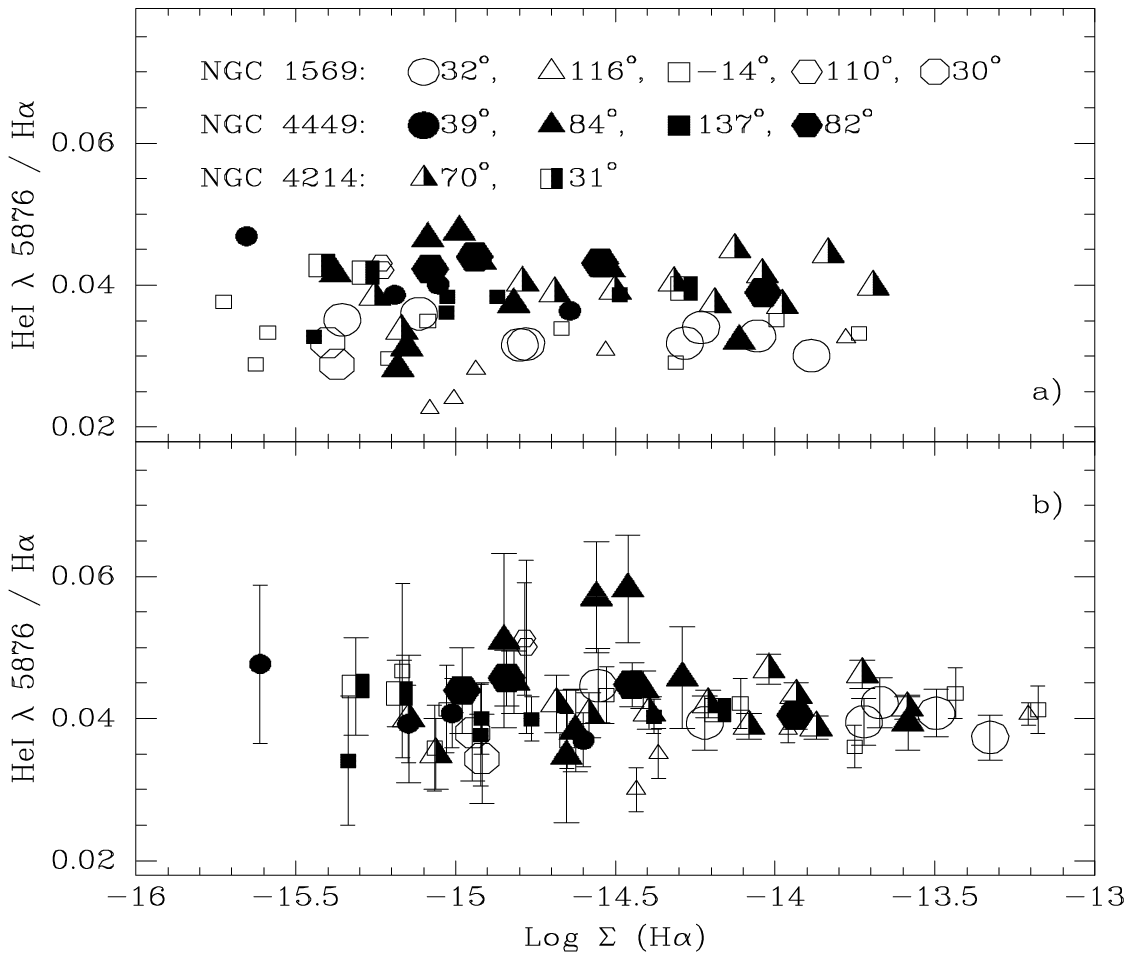}
\caption{
Dependence of the \heha\ intensity ratio on \Ha\ surface brightness:
(a) without corrections for reddening, and
(b) with the de-reddened line ratios and surface brightnesses.
The surface brightness at the positions represented by large symbols 
is in units of \sb.
The small symbols denote non-photometric data and could be misplaced
by as much as $\pm 0.5$~dex in  surface brightness.  
}
\label{fig:data}
\end{figure}

\begin{figure}
\plotone{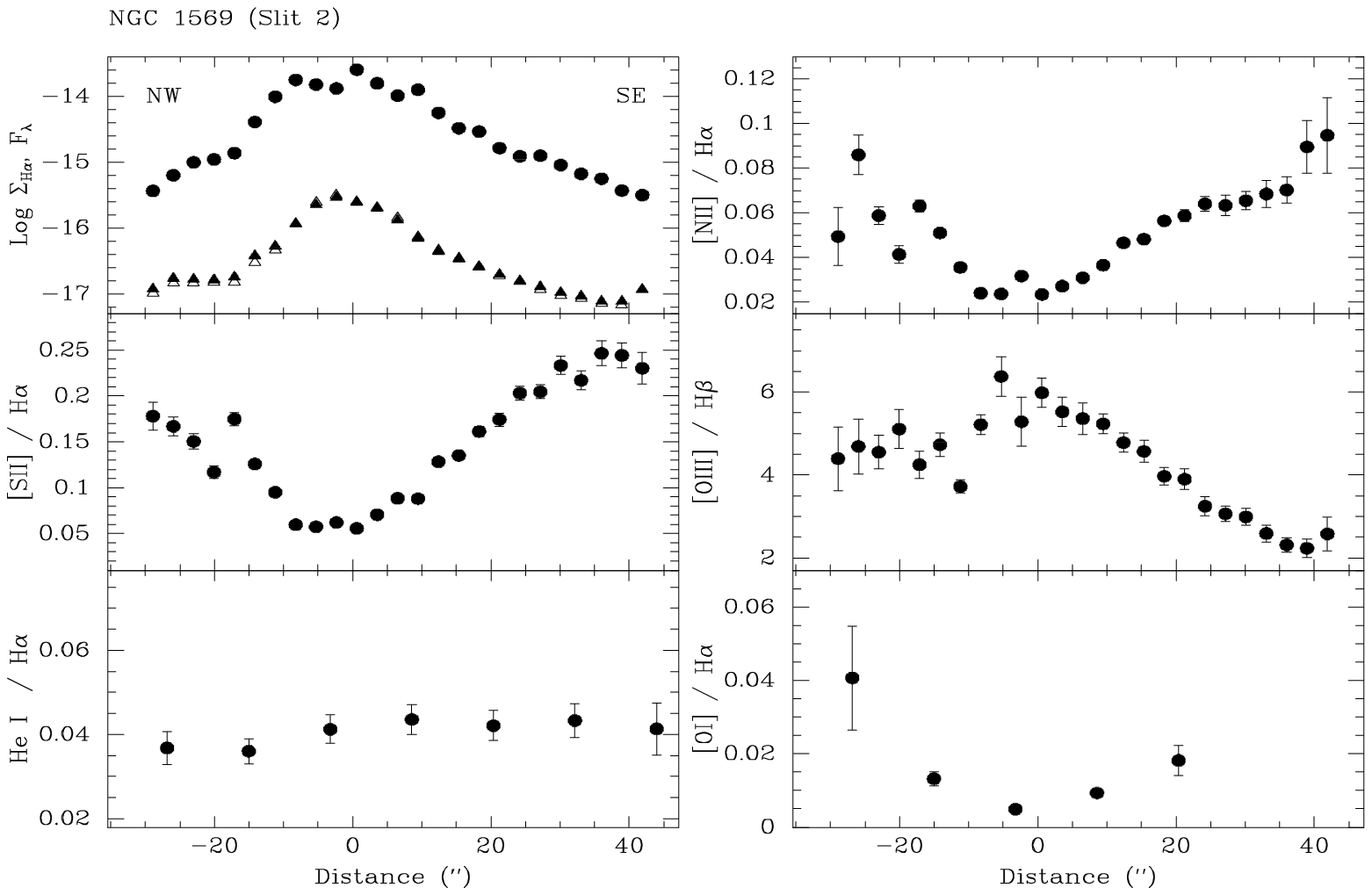}
\caption{
The changing
emission-line spectrum across \n1569, PA = 32.4\deg.
{\em Top Left: }
\Ha and continuum surface brightness versus separation
from the brightest HII region along the slit.
Counterclockwise from top left, the emission-line
ratios are:
\s2ha;
~\heha;
~\N2;
~\o3hb; and
~\O1ha.
} 
\label{fig:slit}
\end{figure}

Our spectra also show evidence for emission from gas excited by 60 -- 100\kms
shocks (Martin 1997a).
These results are discussed in depth in a second paper, and we comment
here only on the  sensitivity of \heha\ to shock velocity.
In the  models of Shull \& McKee (1979),  He~I~\line\ 5876 emission
is negligible until shock speeds reach 80\kms;
even then, the  \heha\ intensity ratio is  only 0.005.
Increasing the shock speed to 100\kms, however, raises \heha\  to 0.047.  
It is possible then that  shocks may contribute to
the high values in \fig~\ref{fig:data}.
However, we suspect the correction is not large because
(1) shock-excited gas typically contributes $\sles\ 20 - 30\%$ of the emission,
and (2) while the relative contribution from shocked gas grows with distance
from the star forming regions, the \heha\ ratio exhibits no
 systematic variation (\fig~\ref{fig:data}).
The relative amounts of He and H ionization in the DIG are therefore a measure
of the hardness of the Lyman continuum radiation ionizing the DIG. 

%\subsection{Constraints on the Stellar Mass Function}
\subsection{The Lyman Continuum}

\fig~\ref{fig:models} illustrates the dependence of the 
 \heha\ intensity ratio of an HII region on the stellar luminosity
of He-ionizing photons relative to H-ionizing photons,  Q(He)/Q(H).
The ratio \heha\ increases linearly with Q(He)/ Q(H)
until H begins to compete for $h\nu > 24.6$~eV photons 
and then saturates at a constant value when 
the volume-averaged $\chi(He) \approx \chi(H) \approx 1$
(Osterbrock 1989).  In \fig~\ref{fig:models}, the 
arrow illustrates the analytic relation for the linear rise; and the
turnover is illustrated by the line ratios of the model nebula ionized
by stars with ${\rm Q(He)/Q(H)} \approx {\rm He/H}$.
Since the mean $\heha\ = 0.041$ in  these
irregular  galaxies, we see from \fig~\ref{fig:models} that the Lyman continuum 
must  have Q(He)/Q(H) greater than 0.12;  and this mean intersects the relation for the
He/H~=~0.85  models at  ${\rm Q(He)/Q(H)} \approx\ 0.25$.  
The \heha\ intensity ratio is, however, not very sensitive to  the
spectral hardness at temperatures hotter  than $T_{eff} = 40,000$~K. 

The equivalent stellar mass depends on the stellar metallicity, the stellar
atmospheres chosen, and the adopted grid of evolutionary models. 
To illustrate the sensitivity  to these assumptions, four mass scales are shown at the 
top of  \fig~\ref{fig:models}.  For example, using the most recent atmospheres from 
Schaerer \et (1996) and the evolutionary models of Schaller \et (1992) as
parameterized by Vacca \et (1996), the spectrum is harder than that of a $\sim 30\msun$
solar metallicity star (scale {\it c}).  If we use the same models as Heiles \et (1996) for
a  direct comparison with their analysis of the Milky Way DIG, the minimum mass star that has a
spectral hardness consistent with the He-ionization of the DIG in the irregular
galaxies is $\approx 44$\msun.

Of course many stars contribute to the ionization of the DIG, and
the  Q(He)/Q(H) ratio of the ensemble will be less than that emerging from
the most massive star. 
The 0.25\zsun\  evolutionary synthesis models of Leitherer \& Heckman (1995), for example,
predict a spectral hardness  Q(He) / Q(H) = 0.23 and Q(He)/Q(H) = 0.07  from 
stellar populations continuously forming stars with a Salpeter initial
mass function (IMF) and upper mass limits of $m_u = 100\msun$ and 30\msun\, respectively.
The ratio can be  considerably higher, Q(He) / Q(H) = 0.32, if the burst is less than a
few~Myr old.
Hence, 30\msun\ is only  a lower limit 
on the upper mass cut-off of the stellar population ionizing the DIG.

%which rely on the Maeder (1990) and Kurucz (1992) models,
%(Compare columns 4 and 7 in Table~4 of Heiles \et (1996) for example.)

\subsubsection{Direct Evidence for Massive Stars}

Optical and ultraviolet spectra provide direct evidence for massive
stars in all  three galaxies. 
Several HII regions in \n4214 show strong,
broad \ion{He}{2} \line 4686 and \ion{C}{4}~\lines 5808 emission lines
from WN and WC stars (Sargent \& Filippenko 1991).  Our spectra also reveal
both of these features in several HII regions in \n4449 as well as
broad \line 4686 in \n1569 (cf. Drissen, Roy, \& Moffat 1991; Gonzalez-Delgado
\et 1997).
These Wolf-Rayet stars  are thought to be the short-lived descendants
of the most massive O stars ($M \ge 35 \msun$)
(Conti \et 1983; Humphreys, Nichols, \& Massey 1985).
Spectral synthesis modeling of the ultraviolet
continuum from  the most prominent cluster in \n4214 suggests  several
hundred O stars are present in addition to the $\sim 30$ Wolf-Rayet stars
(Leitherer \et 1996).
Hence, the hard interstellar Lyman continuum in these irregular galaxies  is not
unexpected. 
It is, rather,   the contrast between the spectral energy distribution of the
photons ionizing the DIG in irregular galaxies and the Milky Way that is of
interest.

% Since \heha\ is not sensitive to the ionizing continuum for
% $Q(He)/Q(H) > 0.12$, we cannot constrain any small changes in the 
% shape of the ionizing continuum.

% at metallicities near solar

\subsection{The Morphology Problem}

The morphology problem in the Galaxy, again,  is how the ionizing photons 
from O and B stars,  scale height $\sim 100$~pc, can reach the DIG when 
their absorption mean free path is only $0.5 (0.1~{\rm cm}^{-3} / n_H)$~pc 
at  1 Rydberg (e.g. Dove \& Shull 1994; Miller \& Cox 1993).
In irregular galaxies, the DIG also extends
over a kiloparsec from the main star forming regions.
And, in both environments, spectra of DIG regions 
$\sim 100$  times fainter than the discrete HII regions
(Reynolds 1991; Reynolds \& Tufte 1995),
indicate  the  ionization parameter is very low
(D\"{o}mgorgen  \& Mathis 1994); Martin 1997a). 
These conditions are consistent with a distant source of ionizing photons.
However, 
the spectroscopic signature of gas photoionized by a distant
association is nearly indistinguishable from that of a dilute HII region;
and it is worthwhile to re-examine
the possibility that the DIG in irregular galaxies might be ionized locally
and the radiation transport problem avoided.

About 80\% of the
ultraviolet light from starburst regions is believed to come from
massive stars between and beyond the young clusters (Meurer \et 1995). 
However, since most studies of the individual massive stars in these galaxies have
focused on star clusters, it is not clear at present how far from the clusters
the young field population might extend (cf. Gallagher \et 1996).
In extreme  environments, such as the outflow extending several
kpc above the disk of \n1569 (Heckman \et 1995), it seems highly unlikely that 
the DIG is ionized  by  nearby stars.  In addition to the difficulties of
forming stars in the tenuous outflow, the dramatic increase 
in \Ha equivalent width with radius 
(cf. \fig~5a Waller 1991) is most naturally explained by  photons escaping the
starburst region and/or shock excitation. 
However, the situation is much less clear a few hundred parsecs to
1.5~kpc away from the clusters -- the type of environment well-sampled
by our spectra; and it is here that we consider whether
the DIG could be comprised solely of low surface brightness HII regions
excited by a young field population

\subsubsection{Is Local Ionization Consistent with the Spectral Gradient?}

Such a question is hard to answer definitively without observing the presence/absence
of an extended population of hot stars.  However, some insight
can be obtained by investigating whether
such an assertion is  compatible with  the observed  spectral gradient.
For the purpose of illustration, we take the viewpoint that all the HII regions
are ionization bounded, and that the radial gradients in \Ha surface brightness
($\Sigma$) and ionization parameter ($U$) result from changes in the HII region
population.  The simplest possible geometry for the HII regions surrounding an
isolated star or cluster is a homogeneous Str\"{o}mgren sphere a fraction
$\epsilon$ of which is filled with gas clouds of density $n$.  
The nebular ionization parameter will scale as 
	\begin{equation}
	U \propto Q^{1/3} n^{1/3} \epsilon^{2/3},
	\end{equation}
where $Q$ is the ionizing luminosity of the star or cluster; and
the average \Ha surface brightness of the nebulae will scale as
	\begin{equation}
	\Sigma \propto Q^{1/3} n^{4/3} \epsilon^{2/3}.
	\end{equation}
These scaling relations are reasonably
robust with respect to the local nebular geometry.  For example, if the circumstellar
medium surrounding each massive star has been swept into a thin shell of thickness
$\bigtriangleup R = (4/3 \pi R^3 n_0) / (4 \pi R^2 4 n_0) = 1/12 R$, the same scaling
arguments would continue to hold under the assumption that the nebulae remain
ionization bounded.  The absolute surface brightness and ionization parameter
of the shell and filled sphere
models would of course differ, but they scale in the same manner with $Q$, $n$, and
$\epsilon$.

From equations~2 and~3, we see that a large-scale gradient
in the ambient gas density is insufficient by itself to  explain the spectral gradient.
Using equation~2,  a smooth change in gas density of a factor of $10^3$
between the  giant HII regions and the DIG could produce
the observed drop, a factor of ten,  in  ionization parameter.
Equation~3 predicts
the surface brightness of the low density nebulae would, however,   be
a factor of $10^4$ times fainter than the giant HII regions --  
a much greater contrast than observed.

In principle, a decrease in the  luminosity of the star clusters with
galactic radius could produce a  gradient in the nebular spectrum
and alleviate the need to transport ionizing photons large distances.
The three order of magnitude drop in $Q$ required to reduce 
the ionization parameter by a factor of ten would be similar to the
difference in ionizing luminosity between a giant HII region and a single hot
star, so individual, isolated  stars would be ionizing the lowest surface
brightness DIG.  
Although the accompanying reduction in \Ha surface brightness is only a factor of ten,
the surface brightness could be decreased further to the observed factor of 100 by 
a mild density gradient in the ambient medium without
decreasing the ionization parameter much beyond the measured range.
Ionization of the DIG by a population of field O and B stars
cannot, therefore, be  ruled out.  

Such an interpretation, however,  implies 
a very smooth radial change in the luminosity of the ionizing star clusters to
generate the smoothness of the spectral gradient.  Until such spatial  changes
in the cluster luminosity function are observed,
we remain highly skeptical of this  explanation
and continue to favor a scenario where the DIG is powered
mainly by the major star forming regions in these irregular galaxies.
The photons ionizing the DIG are most likely leaking out of the giant
HII regions and traveling very large distances before being absorbed.
Additional support for this leakage is provided by
Leitherer \et (1996) who  resolved the central starburst in \n4214
and demonstrated that the HII region  around it is density bounded.

\subsection{Speculation on the Variations among Galaxies} 

Since the He ionization fraction in the DIG of these irregular galaxies
is so much higher than in the Milky Way, one might question whether
the extra-HII region \Ha emission in Magellanic irregular galaxies is a good analogy to
the Reynolds Layer in the Milky Way?
Many properties of the widespread ionized gas in these galaxies are compared to 
those in the Mikly Way DIG in Table~2.
The differences are not limited to the He ionization fraction.
The surface brightness of the regions studied spectroscopically
in the irregular galaxies is  about five times  brighter than  even  the DIG in
the Galactic plane studied by Reynolds.
In addition to diffuse (i.e. unresolved?) emission,
the DIG in the irregular galaxies  contains a highly structured component of shells, arcs, and
radial filaments, sometimes referred to as ``interstellar froth''
(Hunter \& Gallagher 1990). Despite  these differences, we
believe the analogy is   interesting because the DIG in irregular galaxies seems to share 
the morphology problem with the Reynolds Layer.

The \heha\ spectral differences among different types of galaxies demonstrate
that the  physical conditions within the warm-ionized phase of the ISM vary.
The surprisingly low \heha\ ratio  is not limited to 
the Milky Way.  Rand (1996) recently reported $\heha\ \approx 0.034$ 
about 1.5~kpc above the plane of \n891, an edge-on Sbc galaxy with 
a prominent DIG component. The inferred He ionization, about 70\%, is
intermediate to irregular galaxies and the Milky Way.  The high \N2 in \n891
relative to the Milky Way, however, still seems to require a harder spectrum
than the \heha\ ratio (Rand 1996).  

It remains unclear why the He ionization fraction in the DIG of \n891 and
the Milky Way is lower than expected.  Compared to the irregular galaxies, the
dust content and metallicity  are higher; but 
the ionizing luminosity of the largest star-forming complexes tends to be 
smaller than in the irregular galaxies.  
Accounting for this absorption will, however,  harden the Lyman continuum 
thereby magnifying the discrepancy (Sokolowski 1994; Shields \& Kennicutt 1995).  
For the Milky Way, Dove \& Shull (1994) have demonstrated how 
the hierarchical network of Stromgren spheres created by 
the distributions of hot stars and gas
increases the probability of an ionizing photon reaching the DIG.
It is unclear whether a higher escape probability from the vantage point
of large clusters in irregulars could make  the escaping Lyman continuum
relatively harder.  We are reluctant to suggest systematic differences in the
IMF since Massey (1997 and references therein)
measure similar IMFs in the Magellanic Clouds and Milky Way;
however, as more data become available, the possibility of systematic 
variations in  the IMF may need to  be re-considered.

A final, but important, consideration for reconciling the 
interstellar Lyman continua is the  accuracy of the stellar atmosphere calculations.
One of the two massive stars observed with EUVE, $\epsilon $~CMa, is a case
in point. The Lyman and He~I continua of this  B2~II star  are \sgreat\ 30 times 
higher than expected making it the dominant source of ionizing photons within
a few hundred parsecs of the sun (Cassinelli \et 1995).
More reliable predictions of the ionizing flux from stars 
this cool are needed to determine whether B stars can ionize much of the Galactic DIG.
New model atmospheres for O stars including the velocity gradient in the
stellar wind predict an increase in
the He~II continuum by several orders of magnitude (Schaerer \& de Koter 1996), and 
similar effects are expected in the He~I and Lyman continua of B stars
(Najarro \et 1996). 
Since the  objection to ionizing the DIG in the Milky Way with late O and
early B stars is the high total star formation rate implied (Heiles \et 1996),
we explored the effect of the new model atmospheres of Schaerer \& de Koter (1996) 
on the required star formation rate.  For the same cutoff in upper mass
as that used by Heiles \et (1996), the new model atmospheres predict 
a lower star formation rate because the luminosity of hydrogen ionizing photons from
stars with initial masses between 16\msun\ and 45\msun\ is increased.
However, the ionizing continua are also harder, and the mass of the most
massive star
that can contribute to the ionization of the DIG is reduced from 39\msun\
to 25\msun (see \fig~\ref{fig:models}).  With this revised upper mass limit on the ionizing population
of stars, the implied galactic star formation rate is twice that derived
by Heiles \et (1996) making the discrepancy with the measured star formation
rate in the Galaxy (e.g. McKee 1989)  even larger

\subsection{A Consistent Picture in Irregular Galaxies}

We measure a mean $\heha\ \approx $ 0.040, 0.042, and 0.043 across
\n1569, \n4214, \n4449, respectively, and 
 find no evidence for a systematic variation in \heha\ with either
\Ha surface brightness or distance from the nearest giant HII region.
This high relative ionization fraction requires an ionizing continuum at least
as hard as that supplied by a 30\msun\ star, which is completely
consistent with the expected hardness of the radiation from  the young starburst regions
in these galaxies.  Although the transport of these photons over a kiloparsec  
to the DIG is not well-understood,   
it seems plausible that the plethora of expanding supershells of gas
in these irregular galaxies 
(Hunter \& Gallagher 1990; Hunter \& Gallagher 1996; Martin 1996)
creates a rather porous ISM which allows a fraction
of the ionizing photons to travel large distances before absorption.

\acknowledgements{
We thank Joe Shields for  sharing his insight
on He line emission and photoionization codes,
Rich Rand and Carl Heiles
for  discussions about the ionization of the DIG, and 
Rene Walterbos, the referee, for his insightful comments.
We also extend our gratitude to Craig Foltz, Mike Lesser, 
and Gary Schmidt for their contributions to the 
spectrograph and to the MMT telescope operators,
Carol Heller, John McAfee, and Janet Miller, for their
assistance.  Funding for this work was provided by the
National Science Foundation through Grant  AST-9421145.
Additional support was provided by NASA through
Hubble Fellowship grant HF-01083.01-96A.
}

%Gary Ferland for public access to his photoionization
%code, CLOUDY.

%
%TABLES
%

%\input{table1.tex}		% General Properties:  Table~\ref{tab:table1}

\begin{table}
\caption{Properties of the Galaxies  \label{tab:table1}}

\vskip\the\baselineskip\
\begin{tabular}{lllll}
\hline
Galaxy			& \n1569	&	\n4214		& \n4449	& Milky Way	\\
\hline
\hline
d (Mpc)\tablenotemark{a}& 2.2		& 3.6			& 3.6		& \nodata	\\
c(\Hb)\tablenotemark{b}	& $0.85 \pm 0.3$& $0.16 \pm 0.1$        & $0.29 \pm 0.3$ & \nodata      \\
%$\Gamma$  ($s^{-1}$)	& 1.2e53	& 1.0e53		& 1.4e53	& $ \sgreat 1.4e53$	\\
$\Gamma$ ($s^{-1}$)\tablenotemark{c}  	& 3.9e52	& 2.1e52		& 4.5e52	& $ \sgreat$ 1.4e53	\\
  b\tablenotemark{d}	& $\gapp 1$	& $\gapp 1$		& $\gapp 1$	& $\le 1$	\\
$\log $(O/H)\tablenotemark{e}	& $-3.78 \pm 0.07$	& $-3.78 \pm 0.07$		& $-3.69 \pm 0.07$	& -3.07	\\
$M_{\rm HI}$ (\msun)\tablenotemark{f} & 8.4e7		& 1.1e8			& 2.4e9		& 4.8e9	\\
Type\tablenotemark{g}             & IBm                 & IAB(s)m               & IBm           &    \nodata     \\   
\hline
\end{tabular}

\tablenotetext{a}
{Adopted distance (Martin 1996).} 

\tablenotetext{b}
{Logarithmic extinction at \Hb (see text).}

\tablenotetext{c}
{Recombination rates calculated 
from the \Ha fluxes  of Kennicutt \& Kent (1983) assuming an electron
temperature of  $10^4$~K  and the extinction in row~2.  
The Galactic rate is discussed by Heiles \et (1996).}

\tablenotetext{d}
{Stellar birthrate parameter, the ratio of the current star formation rate to the past
average rate (Kennicutt  1983; Kennicutt \et 1994).}

\tablenotetext{e}
{Oxygen abundances for irregular galaxies from Martin (1997a); Galactic value from
Grevesse and Anders (1989).}

\tablenotetext{f}
{Mass of neutral hydrogen from
Reakes (1980), Hunter, Gallagher, \& Rautenkranz (1982), Hunter \& Gallagher (1986), and 
Kulkarni \& Heiles (1987).}

\tablenotetext{g} 
{Third Reference Catalog of Bright Galaxies}

\end{table}

\begin{table}
%\scriptsize
\ssp

\caption{The DIG in Different Types of Galaxies   \label{tab:tabR}}

\begin{tabular}{lllll}
\hline
Property	&	Milky Way  	& \n1569	& \n4449	& \n4214 \\
\hline
\hline
\heha \tablenotemark{a}	&	\nodata		& 0.040		& 0.043		& 0.042		\\
\heha \tablenotemark{b}	& $\sles 0.011$		& 0.032		& 0.039		& 0.040		\\
\xhe \tablenotemark{c}		& $\sles\ 0.23$			& $\sim 1$	& $\sim 1$	&	$\sim 1$	\\
				& $< 0.13 $			& 		&		&		\\
$\Sigma_{\Ha}^{\rm DIG} $\tablenotemark{d} (R) & 1 - 10	& 100 		& 50 	& 50 	\\
$\Sigma_{\Ha}^{\rm HII}$ (R)				& 10 - 500	& 10,000	& 4,500	&  4,500	\\
$\log U$ (DIG)\tablenotemark{e} & $\sles -4.1$	& -3.9		& -4.3		& -3.9		\\
$\log U$ (HII)			& \nodata	&  -2.24	& -3.2		& -2.9	\\
%Distance\tablenotemark{f}	(pc)	&$\approx 1000$	& 950 (400) 	& 1900 (1000)	& 1200 (800)	\\		
Distance\tablenotemark{f}		&$\approx 1000$	& 950 		& 1900 		& 1200 	\\		
					&		& (400)		& (1000)	& (800)	\\		
Morphology\tablenotemark{g}		& Diffuse	& Diffuse +	& Diffuse +	& Diffuse +	\\
			&		& Filaments +	& Filaments	& Filaments 	\\
			&		& Wind		&		& 	 	\\
$M_u$  (M$_\odot)\tablenotemark{h}$	& $\le 39$	&	$> 30$	&	$>30$	& $> 30$	\\
\hline
\end{tabular}

\vskip\the\baselineskip\
Notes --
(a) Reddening corrected emission-line ratio.
(b) Reddened line-ratio.  Local value from Reynolds \& Tufte (1995).
(c) Relative ionization fraction.  Galactic values from Reynolds \& Tufte (1995) and
Heiles \et (1996).
(d ) The \Ha surface brightness 
of the DIG  compared to that of
HII regions in the Im's and  discrete sources in the Galactic plane
(Reynolds 1983, 1984).
One Rayleigh (R)  is $10^6$ photons  s$^{-1}$ cm$^{-2}$ per $4 \pi$ sr.

(e) Logarithm of ionization parameter, $U$, defined 
in terms of the ratio of ionizing photons to matter
at the Str\"{o}mgren radius, $R_S $, in the absence of absorption
such that $U\equiv \frac{Q}{4 \pi R_S^2 n c}$ 
(Martin 1997a;
Domg\"{o}rgen \& Mathis 1994).
(f) Projected distance from giant HII regions to the most
extended \Ha emission in our images 
and, in parentheses, to the boundary of our He~I  measurements.
Estimated scale height of Galactic diffuse, ionized gas (Reynolds 1991).  
(g) Morphology of extra-HII region ionized gas.
(h) Upper-mass limits  on present day mass function
of stars contributing to the ionization of the DIG (this paper and
 Heiles \et 1996).
\end{table}

\newpage

%
% REFERENCES
%

\newpage
%
% FIGURES
%
%\begin{center}
%FIGURE CAPTIONS
%\end{center}
%
\begin{figure}
\caption{
{\it Bottom:}
Theoretical dependence of \heha\  on the hardness of 
the stellar  continuum.  The  circles denote the line 
intensities of  nebulae  ionized by
Kurucz model atmospheres (1979) with effective temperatures of 35,000~K, 40,000~K,
45,000~K, and 50,000~K 	(CLOUDY version 84.09, Ferland 1993).
The clouds have ${\rm O/H} = 0.20 \ohsun$;  and the closed and open circles represent models
with He/H abundance ratios of 0.08 and 0.09, respectively.
The  \heha\ ratios observed in the DIG are indicated. 
{\it Top:}
The  mass of a main sequence star producing an ionzing spectrum with a given hardness.
The four mass scales were derived from:
(a)  the stellar evolution grid of
Maeder (1990) at the Kurucz effective temperature and solar 
metallicity  (0.25\zsun\ in parentheses).
(b) the same T(M,Z) as in {\it a}, but
the stellar atmospheres of Schaerer \et (1996) which include  wind
effects and produce a harder ionizing spectrum at a given temperature,
(c) the same stellar atmospheres as {\it b}, but using the evolutionary mass  from
Vacca \et (1995) as interpolated from the grid of solar metallicity stellar evolution
calculations by Schaller \et (1992), and
(d) for comparison, from equation~10 of Heiles \et (1996) which relies on older 
atmospheric models.
}
%, and the thin line shows $He/H = 0.1$ for comparison.
%The ratio $Q(He) / Q(H)$  represents the relative number of photons
%with $h\nu > 24.6$~eV to photons with $h\nu > 13.6$~eV from the Kurucz
%model atmospheres.
%(a) The mass derived from the T(M,Z) relation defined by the stellar evolution grid of
%Maeder (1990) at the Kurucz effective temperature and a metallicity 0.25\zsun (1.0\zsun
%in parentheses).
%(b) The stellar atmospheres of Schaerer \et (1996) which include non-LTE effects, wind
%effects, and line blanketing produce a harder ionizing spectrum at a given temperature.
%The same T(M,Z) as in ``a.''
%(c) Same stellar atmospheres as ``b,'' but the evolutionary mass is taken from
%Vacca \et (1995) as interpolated from the grid of solar metallicity stellar evolution
%calculations by Schaller (1992).
%(d) For comparison, the mass vs Q(He)/Q(H) relation predicted by equation~10 of Heiles \et (1996).
\label{fig:models}
\end{figure}

\begin{figure}
\plotone{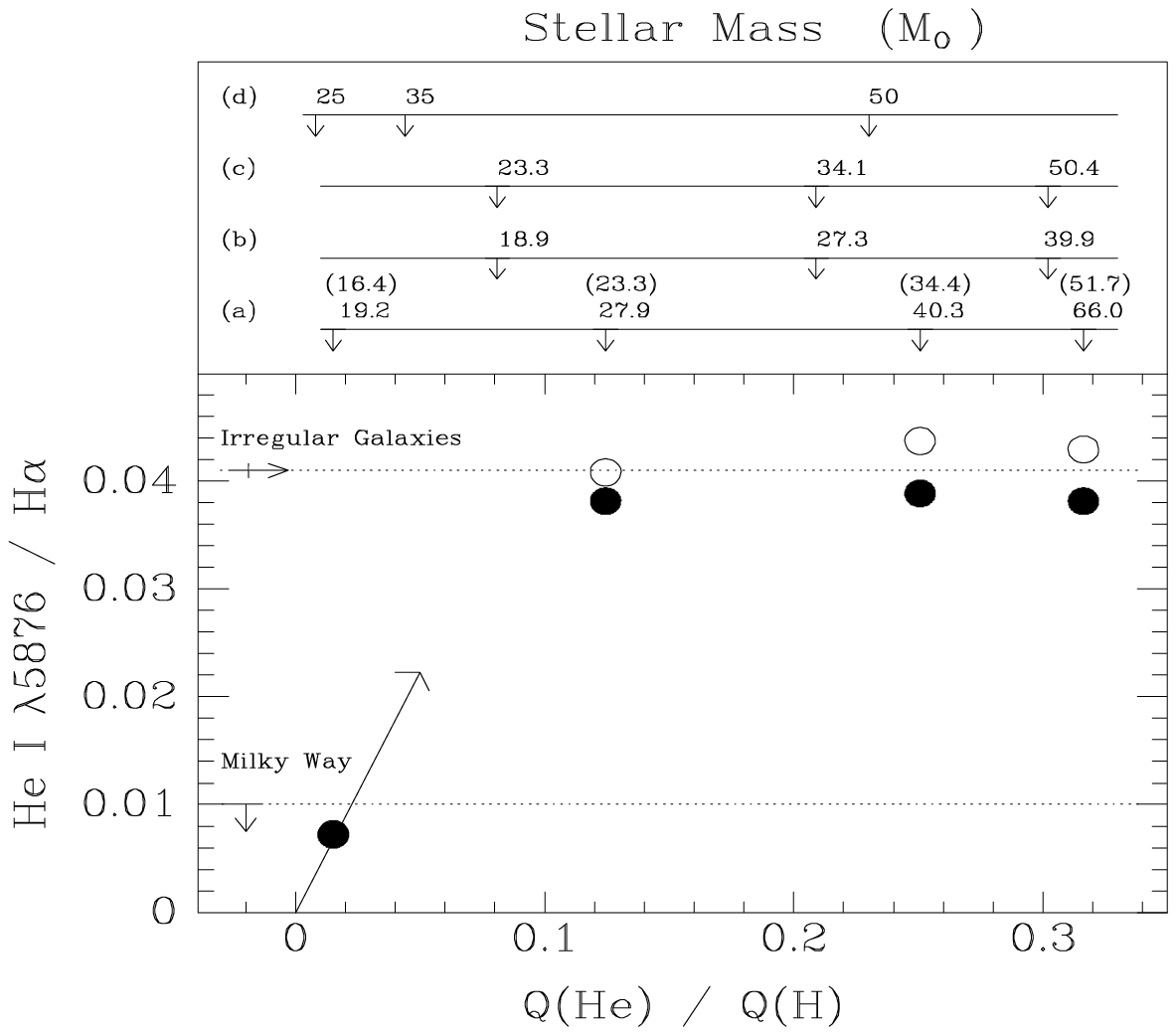}
\end{figure}

\end{document}